\documentclass[preprint,review,12pt]{elsarticle}

\usepackage{amssymb}

\journal{Carbon}

\begin{document}

\begin{frontmatter}

\title{Perfect valley filter controlled by Fermi velocity modulation in graphene}

\author[1]{A. R. S. Lins}
\author[1,2]{Jonas R. F. Lima\corref{cor1}}
\cortext[cor1]{}
\ead{jonas.lima@kit.edu}
\address[1]{Universidade Federal Rural de Pernambuco, Departamento de Física, 52171–900, Recife, PE, Brazil.}
\address[2]{Institute of Nanotechnology, Karlsruhe Institute of Technology, D-76021 Karlsruhe, Germany}

\date{\today}

\begin{abstract}
In this work we investigate the effects of a Fermi velocity modulation in a valley filter in graphene created by a combination of a magnetic and electric barrier. With the effective Dirac equation of the system, we use the transfer matrix formalism to obtain the transmittance. We verify that the valley transport in graphene is very sensitive to a Fermi velocity modulation, which is able to choose which valley will be filtered with perfect filtering. Also, it is possible to use a Fermi velocity modulation to filter both valleys or to make the valley filter transparent. It reveals that the Fermi velocity is a powerful tool that can be used to tune a graphene valley filter, since it has a total control in its transport properties.
\end{abstract}

\end{frontmatter}

\section{Introduction}

Since its first experimental realization in 2004 \cite{Novoselov}, graphene has attracted a great deal of attention due to both, its conection between different branchs of physics and its potential of application \cite{RevModPhys.81.109}. One of the most interesting features of graphene is that its low-energy electronic excitations can be described by a Lorentz invariant theory \cite{Valeri}, in contrast to usual semiconductors. It is due to the existence of two independent Dirac points in the electronic structure of graphene that appear in the points $K$ and $K^{\prime}$ in the Brillouin zone, which characterizes the two valleys in graphene. As a consequence, due to the Klein tunneling \cite{Katsnelson}, electrostatic potential barriers are invisible to quasiparticles with normal incidence, which limits the use of graphene in electronic devices. The electronic confinement in graphene can be improved, for instance, by openning an energy gap in its electronic structure \cite{SiC,bdg2,BN,doped,bdg} or including magnetic barriers \cite{PhysRevB.73.241403,PhysRevLett.98.066802,DEMARTINO2007547}. 

In 2007, a seminal work proposed a way of occupying a single valley in graphene, producing a valley polarization \cite{Rycerz2007}. The proposed valley filter should occur in a ballistic contact point with zigzag edges. Due to the large momentum separation between the two valleys in graphene, valley information could be preserved for a long distance \cite{PhysRevLett.100.056802}. It attracted a great deal of attention in the investigation of the use of graphene in the valleytronics. Different ways of generating a valley polarized current in graphene were proposed, such as with electromagnetic fields \cite{PhysRevB.91.115422,doi:10.1063/1.4866591,PhysRevB.85.155415,Wang2017}, trigonal warping \cite{PhysRevLett.100.236801,Pereira_Jr_2008}, line defects \cite{PhysRevLett.106.136806,PhysRevB.87.195445,PhysRevB.89.121407}, lattice strain \cite{PhysRevLett.103.046801,doi:10.1063/1.3473725,doi:10.1021/nl1018063,PhysRevB.88.195426,PhysRevLett.106.176802,PhysRevLett.110.046601,PhysRevLett.117.247702,RIS_0} and also optical fields \cite{PhysRevB.84.195408,PhysRevB.91.041404}. More recently in 2014, the first experimental observation of a valley current in graphene was performed \cite{Gorbachev448}. In 2015, a valley current was also observed in a bilayer graphene \cite{RIS_01,RIS_02}. These observations attracted even more attention in the investigations of how to create and manipulate a polarized valley current in graphene \cite{YU2016451,doi:10.1063/1.4943237,doi:10.1063/1.4967977,ISLAM2016304,ZHANG2018183,Komatsueaaq0194}.

In the last years, various studies have revealed that the electronic and transport properties of graphene can be controlled by a Fermi velocity engineering \cite{LIMA20151372,Lima2017}. For instance, it was obtained that a Fermi velocity modulation in graphene can control the energy gap \cite{LIMA2015179} and also induce an indirect energy gap in monolayer \cite{LIMA201582} and bilayer \cite{Cheraghchi_2013} graphene. The Fermi velocity can also be used to create electrons guides in graphene \cite{Polini,Yuan}, to control the Fano factor \cite{LIMA2018105} and to tune the electrons transmittance from 0 to 1 \cite{doi:10.1063/1.4953865}, which means that it can turn on/off the transport in graphene. The Fermi velocity in graphene can be engeneered, for instance, by the substrate \cite{Hwang}, by doping \cite{doi:10.1002/pssb.200982335} and by strain \cite{PhysRevB.84.195404,JANG2014139}. As the Fermi velocity in graphene depends on the electron concentration \cite{Valeri,Elias,PhysRevB.83.041405}, it is possible to induce a position-dependent Fermi velocity placing metallic planes close to the graphene layer, since the presence of the planes will change the electron concentration in different regions \cite{Polini,Yuan}. Fermi velocities as high as $3\times 10^6$~m/s were already obtained in graphene by electron's concentration modifications \cite{Elias}. However, as far as we know, there are no studies about the effects of a Fermi velocity modulation in the valley polarization of graphene.

Motivated by these studies, in this work we investigate the influence of a Fermi velocity modulation in the transport properties of the valleys in graphene. We consider that the valley-dependent transport properties are generated by a combination of a magnetic and electric barrier. Within the continuum limit, which is based on an effective Dirac Hamiltonian, we use the transfer matrix formalism to obtain the transmittance of the system. We verified that the valley transport in graphene is very sensitive to the modulation of the Fermi velocity. In fact, we obtained that the Fermi velocity has a complete control over the valley transmission through the barriers, being possible, for instance, to choose which valley is transmitted inducing a transmittance equal to 1 for one valley and 0 for the other, and also to induce a transmittance equal to 1 or 0 for both valleys. These results can be used for the fabrication of a graphene-based valley filter, which creates valley polarized currents.

The paper is organized as follows. In Sec. 2 we describe the system and write out the effective Dirac equation. We solve it for the wave function and use the transfer matrix formalism to obtain the transmittance of the system. With the transmittance, in Sec. 3 we numerically obtain and discuss the influence of the Fermi velocity modulation in the valley transport in graphene. The paper is summarized and concluded in Sec. 4.

\section{Model and Formalism}

\subsection{Dirac Equation}

\begin{figure}[hpt]
\centering
\includegraphics[width=0.9\linewidth]{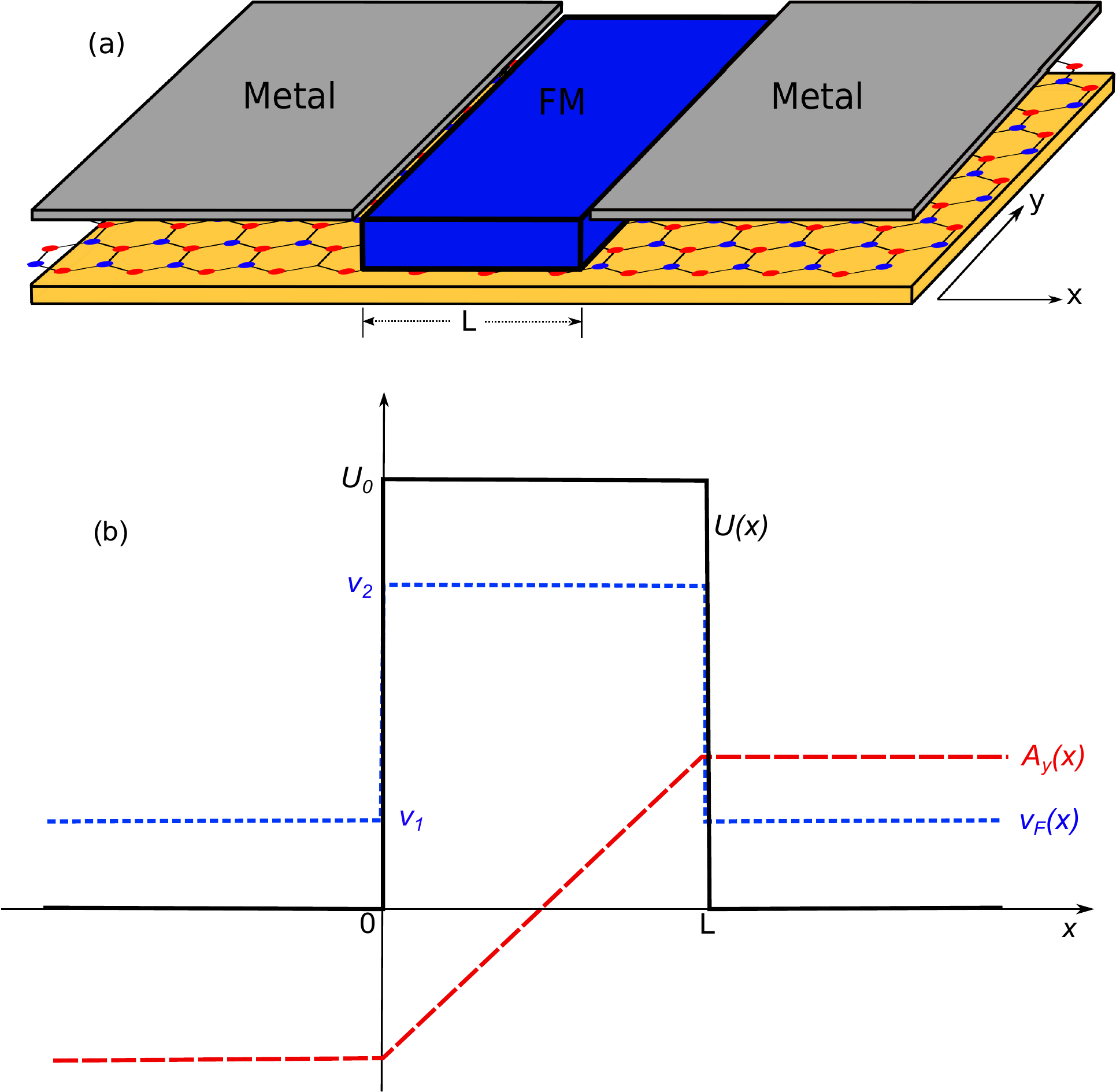}	
\caption{The schematic diagram of the graphene valley filter created by a magnetic and electrostatic barrier of width $L$, with a modulated Fermi velocity. The vector and scalar potential are shown in $(b)$. The Fermi velocity modulation in induced by metallic planes close to the graphene, which will change the electrons concentration in different regions of graphene.}
\label{graphene}
\end{figure}

We are interested here in analize the effects of a Fermi velocty modulation in the valley transport in a graphene layer. As was pointed out in Ref. \cite{PhysRevB.85.155415}, a valley-polarized current can be created by a magnetic barrier plus an electric or energy gap barrier in graphene. Here, we will consider a constant energy gap given by $2\Delta$ induced by the substrate that can be, for instance, SiC \cite{SiC}, so the valley polarization will be the result of a combination of a magnetic and electric barrier. A schematic diagram of the system can be seem in the top of Fig. \ref{graphene}. We consider that the magnetic barrier is created by a magnetic field perpendicular to the graphene sheet, $\vec{B} = B_0 \hat{e}_z$, which is translationally invariant in the $y$ direction, i. e., $B(x,y) = B(x)$. This magnetic field can be created, for instance, depositing ferromagnetic metal (FM) on the top of the layer. The electrostatic potential $U$ also varies only in the $x$ direction and is induced by the FM gate. They are given by
\begin{equation}
B(x) = B_0 \Theta(x) \Theta(L-x),
\end{equation}
\begin{equation}
U(x) = U_0 \Theta(x) \Theta(L-x),
\end{equation}
where $\Theta(x)$ is the Heaviside function. In the Landau gauge, the vector potential is given by $\vec{A} = (0, A_y(x), 0)$, where 
\begin{equation}
A_y(x)=\left\{ \begin{array}{ll}
-B_0 L/2, & x \in (-\infty,0] \\
B_0 x - B_0 L/2, & x \in [0,L] \\
B_0 L/2, & x \in [L,\infty).
\end{array}\right.
\end{equation}
The scalar and vector potential can be seen in the bottom of Fig. \ref{graphene}. We are considering that the Fermi velocity modulation is generated by metallic planes placed near the graphene sheet, as is shown in the top of Fig. \ref{graphene}, which induce
\begin{equation}
v_F(x) = \left\{ 
\begin{array}{ll}
v_1, & x < 0, x > L \\
v_2, & 0\leq x \leq L.
\end{array} \right.
\end{equation}
Since the injection of valley-polarized current can generate a transverse voltage in a graphene with broken inversion symmetry \cite{PhysRevLett.99.236809}, the experimental verification of our results can be performed by measuring how this transverse voltage in the outgoing region can be tuned when the distance between the metallic planes and the graphene changes. 

The effective Dirac equation for the system is given by $H\psi_{\tau} = E \psi_{\tau}$, where
\begin{eqnarray}
H = \sqrt{v_F(x)}\vec{\sigma}\cdot \left(\vec{P} + \frac{e}{c}\vec{A}\right)\sqrt{v_F(x)}+ \tau \Delta \sigma_z + U \sigma_0 .
\label{H}
\end{eqnarray}
Here, $\tau = \pm 1$ labels the two valleys in graphene, $K$ and $K^{\prime}$, $\psi_{\tau} = (\psi_{\tau A}, \psi_{\tau B})^T$ is a spinor that represents the two graphene sublattices, $\vec{\sigma} = (\sigma_x, \sigma_y, \sigma_z)$ is the Pauli matrix acting in the pseudospin of graphene and $\sigma_0$ is the unit matrix. It is important to mention that, due to the position dependence of the Fermi velocity, the Hamiltonian (\ref{H}) had to be modified in relation to its usual form in order to becomes Hermitian \cite{peres}. 

Since the wave functions are translationally invariants in the $y$ direction, we can write $\psi_{\tau}(x,y) = \psi_{\tau}(x)e^{ik_yy}$. Additionally, defining $\phi_{\tau}(x) = \sqrt{v_F(x)}\psi_{\tau}(x)$, the Dirac equation becomes
\begin{eqnarray}
\left[-i\partial_x \sigma_x + (k_y + A_y)\sigma_y \right]\phi_{\tau}(x) = \frac{1}{v_F(x) \hbar} \left(E-\tau \Delta \sigma_z - U \right)\phi_{\tau}(x).
\label{eq}
\end{eqnarray}

In what follow, we introduce dimensionless units, where all quantities will be expressed in units of $B_0$, the magnitude of the magnetic field, and $\ell_B = \sqrt{\hbar c/eB_0}$, the associated magnetic length. Then, $A(x)$ will be written in units of $B_0\ell_B$, $x$ in units of $\ell_B$, $k_y$ in units of $\ell_B^{-1}$ and $E$ in units of $\hbar v_F / \ell_B$. Since the energy scale depends on the Fermi velocity, in this dimensionless units the modulation of the Fermi velocity will be incorporated in the problem by the energy. So, in the incoming and outgoing regions we consider that the electrons have energy $E$, while in the barrier region, they have energy $E\zeta$, with $\zeta = v_2/v_1$.

\subsection{Wave Functions}

The Dirac equation (\ref{eq}) gives rise to two coupled equations given by
\begin{equation}
-i[\partial_x + (k_y + A_y)]\phi_{\tau B}(x) = (E-U-\tau\Delta)\phi_{\tau A}
\end{equation}
and
\begin{equation}
-i[\partial_x - (k_y + A_y)]\phi_{\tau A}(x) = (E-U+\tau\Delta)\phi_{\tau B}.
\label{ceq}
\end{equation}
Uncoupling these equations for $\phi_{\tau A}$, one obtains that
\begin{equation}
\partial_{x}^2\phi_{\tau A}+k^2_x (x) \phi_{\tau A}=0,
\label{eqA}
\end{equation}
where $k_x(x)=\sqrt{(E-U)^2 - (\tau\Delta)^2 - \partial_{x}A_y-(k_{y} + A_y)^2}$.

In regions I and III, $k_x(x)$ is constant. So, the solution of Eq. (\ref{eqA}) is of the form
$\phi^i_{\tau A} = A_i e^{ik^i_x x} + B_i e^{-ik^i_x x}$, where $k_x^i = [E^2 - (\tau\Delta)^2 -(k_{y} + A^i_y)^2]^{1/2}$ and $i = I, III$. Replacing this solution in Eq. (\ref{ceq}), one can obtain that the wave function in these regions can be written as
\begin{equation} 
\phi^i_{\tau}(x)= \Omega_i \left(\begin{array}{cc}A_i \\ B_i \end{array}\right) 
\end{equation}
where
\begin{equation}
\Omega_i(x) = \left(\begin{array}{cc}e^{ik^i_{x}x} & e^{-ik^i_{x}x} \\ \frac{k^i_{x}+ i\left(k_{y} + A^i_y\right)}{(E+\tau\Delta)}e^{ik^i_{x}x} & \frac{-k^i_{x}+ i\left(k_{y} + A^i_y \right)}{(E+\tau\Delta)}e^{-ik^i_{x}x} \end{array}\right).
\end{equation}

At the same way, the solution in region II can be obtained as
\begin{equation} 
\phi^{II}_{\tau}(x)= \Omega_{II} \left(\begin{array}{cc}A_{II} \\ B_{II} \end{array}\right),
\end{equation}
where
\begin{equation}
 \Omega_{II}(x)= \left(\begin{array}{cc}D_{p-1}(q) & D_{p-1}(-q)  \\ 
 \\ \frac{i\sqrt{2B_0}}{E\zeta-U-\tau\Delta} D_{p}(q) & -\frac{i\sqrt{2B_0}}{E\zeta - U-\tau\Delta} D_{p}(-q) \end{array}\right).
\end{equation}
Here,
\begin{equation}
p=\frac{(E\zeta-U)^2-(\tau\Delta)^2}{2 B_0},
\end{equation}
\begin{equation}
q=\sqrt{\frac{2}{B_0}} \left(k_{y}-A^{II}_y\right) 
\end{equation}
and $D_p(q)$ is the parabolic cylinder function.

It is important to remember that $\phi_{\tau}$ is not the wave function of the system. In fact, the wave function $\psi_{\tau}$ is equal to $\phi_{\tau}/\sqrt{v_F}$.

\subsection{Transfer Matrix}

In order to obtain the transmittance of the system, we will use the transfer matrix formalism. In this approach, a matrix 
\begin{equation}
\hat{M} = \left(\begin{array}{cc}M_{11} & M_{12}  \\ 
 \\ M_{21} & M_{22} \end{array}\right),
\end{equation}
called the transfer matrix, relates the wave function on the incoming region to the wave function on the outgoing region. This matrix will be obtained by considering the continuity of the wave function.

Since the Fermi velocity in regions I and III are the same, the transfer matrix that connects $\psi_{\tau}$ in regions I and III is the same that connects $\phi^I_{\tau}$ and $\phi^{III}_{\tau}$. So, the continuity condition of $\phi_{\tau}$ in $x=0$ and $x=L$ gives
\begin{equation}
\left(\begin{array}{cc}A_I \\ B_I \end{array}\right) =\Omega_{I}^{-1}(0) \Omega_{II}(0)\left(\begin{array}{cc}A_{II} \\ B_{II} \end{array}\right)
\label{0}
\end{equation}
and
\begin{equation}
\left(\begin{array}{cc} A_{II} \\ B_{II} \end{array}\right) =\Omega_{II}^{-1}(L) \Omega_{III}(L)\left(\begin{array}{cc}A_{III} \\ B_{III}=0 \end{array}\right),
\label{L}
\end{equation}
respectively. Replacing (\ref{L}) in (\ref{0}), one obtains that
\begin{equation}
 \left(\begin{array}{cc}A_I \\ B_I \end{array}\right) =\hat{M} \left(\begin{array}{cc}A_{III} \\ 0 \end{array}\right),
\end{equation} 
where
\begin{equation}
\hat{M} = \Omega_{I}^{-1}(0) \Omega_{II}(0)\Omega_{II}^{-1}(L) \Omega_{III}(L).
\end{equation}
Then, the transmittance is given by
\begin{equation}
T_{K(K^{\prime})} = \frac{k_x^{III}}{k_x^I} \left|\frac{A_{III}}{A_I}\right|^2 = \frac{k_x^{III}}{k_x^I}\frac{1}{|M_{11}|^2},
\end{equation}
where the factor $k^{III}_x/k_x^I$ was included to ensure current conservation, since the potential vector is different in regions I and III. 

With the transmittance of the system, we can now investigate the influence of the Fermi velocity modulation in the valley transport in graphene. The total conductance of the system at zero temperature can be obtained via the Landauer-Büttiker formula, given by
\begin{equation}
G_{K(K^{\prime})} = G_0 \int_{-\pi/2}^{\pi/2} T_{K(K^{\prime})} \cos \theta_0 d\theta_0, 
\end{equation}
where $G_0 = 2e^2 E L_y /(\pi \hbar)$. $L_y$ is the sample size in the $y$ direction. We also define the efficiency and the valley polarization of the filter as
\begin{equation}
\eta_{K(K^{\prime})} = \frac{T_{K(K^{\prime})}}{T_K + T_{K^{\prime}}}
\end{equation}
and
\begin{equation}
P=\frac{G_K-G_{K^{\prime}}}{G_K+G_{K^{\prime}}},
\end{equation}
respectively.

It is important to point out one characteristic of the system that is induced by the magnetic barrier. The wave vectors written in terms of the incident and emergent angles, $\theta_0$ and $\theta_e$, respectively, are given by
\begin{equation}
k_x^{I} = \sqrt{E^2 - (\tau \Delta)^2} \cos \theta_0, \; k_y = \sqrt{E^2 - (\tau \Delta)^2} \sin \theta_0 + \frac{B_0 L}{2}
\end{equation}
\begin{equation}
k_x^{III} = \sqrt{E^2 - (\tau \Delta)^2} \cos \theta_e, \; k_y = \sqrt{E^2 - (\tau \Delta)^2} \sin \theta_e - \frac{B_0 L}{2}
\end{equation}
The conservation of $k_y$ implies that 
\begin{equation}
\sin \theta_0 + \frac{B_0 L}{\sqrt{E^2 - (\tau \Delta)^2}} = \sin \theta_e,
\end{equation}
which means that there will be transmission through the barrier only if
\begin{equation}
\left| \sin \theta_0 + \frac{B_0 L}{\sqrt{E^2 - (\tau \Delta)^2}} \right| \leq 1.
\label{sin}
\end{equation}
This condition restricts the transmission for a smaller range of $\theta_0$ compared with others barriers. This range decreases as $B_0$ or $L$ increases, and increases as $E$ increases. In special, if $|B_0L/\sqrt{E^2 - (\tau \Delta)^2}| > 2$, all electrons are completely reflected by the barrier.

\section{Numerical Results and Discussions}

\begin{figure}[hpt]
\centering
\includegraphics[width=\linewidth]{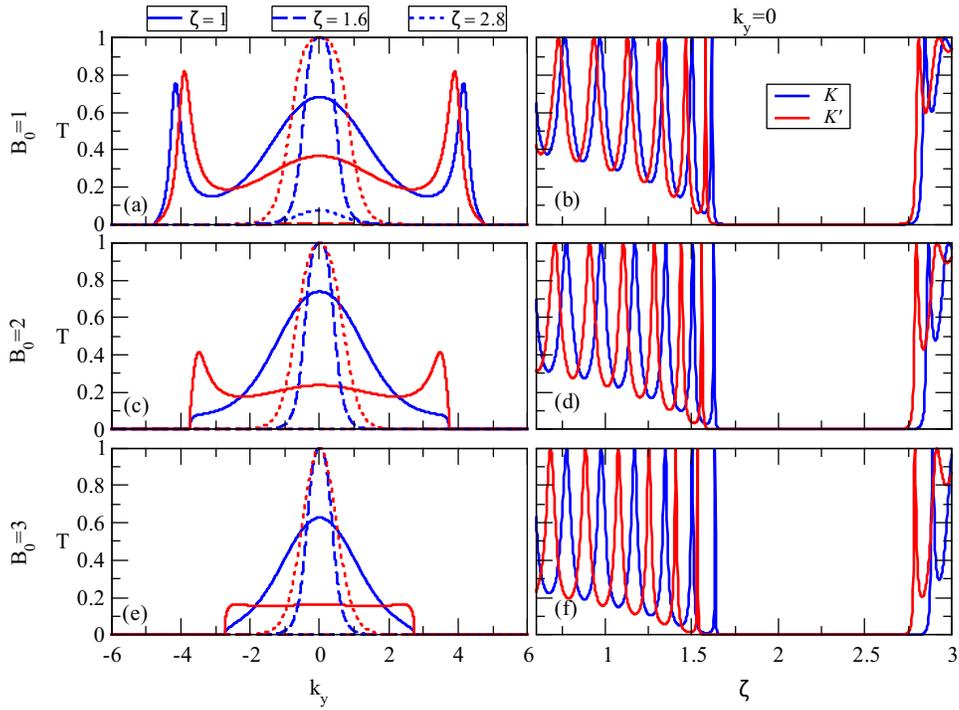}	
\caption{The transmittance for both valleys as a function of the transverse wave vector $k_y$ (left panels) and $\zeta$ (right panels) for different values of $B$ with a fixed value of $E$. We consider here $L=2$, $U_0 = 15.5$, $\Delta = 4$ and $E=7$. The others parameters are depicted in the figure.}\label{b}
\end{figure}

Let us now analyze the effects of the Fermi velocity modulation. In Fig. \ref{b} we plotted the transmittance for various values of $B_0$ with a fixed energy. In Figs. \ref{b} $(a)$, $(c)$ and $(e)$ we consider $T$ in terms of $k_y$ for different values of $\zeta$. As expected, the range of $k_y$ with transmittance different from zero decreases as $B_0$ increases. As can be seen, the difference between the blue and red lines reveals a valley polarized current. In the continuum lines, we have the case of a constant Fermi velocity. Looking at $k_y = 0$, we can note that the Fermi velocity modulation can be used to improve the valley polarized current, as can be seen in the dashed lines, where the transmittance for the valley $K$ can reach 1 and for the valley $K^{\prime}$ can reach 0 as $B_0$ increases, creating an perfect valley filter. Also, the dotted lines reveals that the Fermi velocity engineering can switch the valley polarization. Therefore, the Fermi velocity can substantially improve the valley polarized current in graphene and also control which valley will be transmitted through the barrier. It is important to mention that the improvement in the valley polarization induced by the Fermi velocity can also be obtained for a different value of $k_y$.

\begin{figure}[hpt]
\centering
\includegraphics[width=\linewidth]{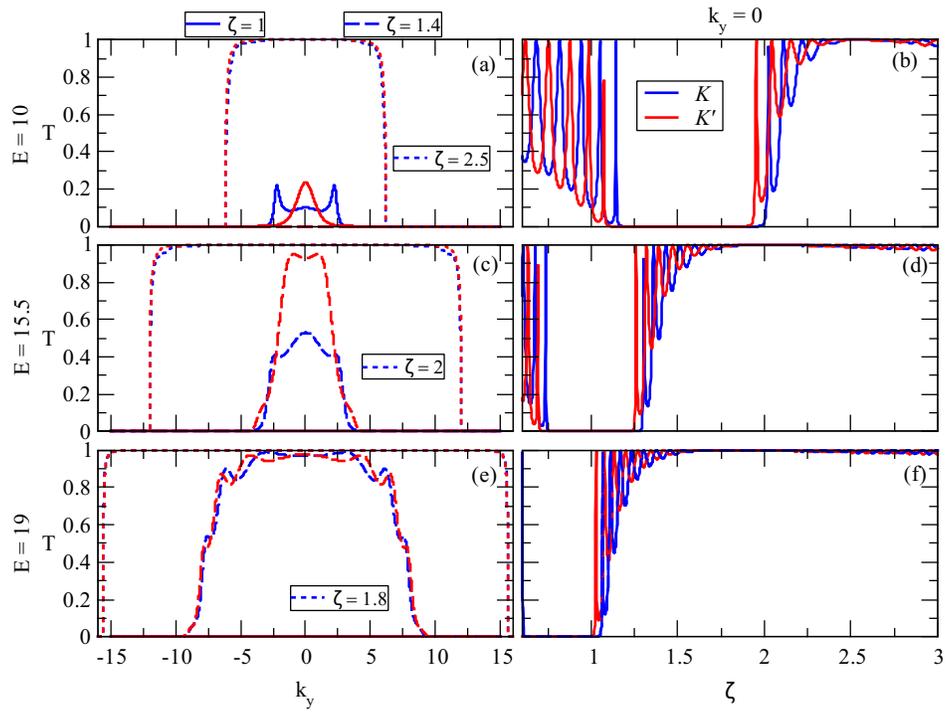}	
\caption{The transmittance for both valleys as a function of the transverse wave vector $k_y$ (left panels) and $\zeta$ (right panels) for different values of $E$ with a fixed value of $B$. We consider here $L=2$, $U_0 = 15.5$, $\Delta = 4$ and $B=3$. The others parameters are depicted in the figure.}
\label{E}
\end{figure}

The transmittance as a function of $\zeta$ for $k_y = 0$ in plotted in Figs. \ref{b} $(b)$, $(d)$ and $(f)$, showing how the valley polarization changes with the Fermi velocity modulation. One can see here three kinds of transport process. The first one is the interband process (from conductance bands to valence bands), which is achieved when $\zeta < (U_0-\Delta)/E$. Here we have the appearance of Fabry-Pérot resonances, which gives rise to the sharp oscillations in the transmission. A valley splitting of these resonances can be observed as a consequence of the magnetic field. As $B_0$ increases, the difference between the resonant peaks for each valley increases. The second one is a tunneling process (through the energy gap), which occurs when $(U_0-\Delta)/E<\zeta < (U_0+\Delta)/E$. In this region, the propagating incident mode becomes an evanescent mode in the barrier region, which can only exist near the boundary. So, the transmission exponentially decays with the distance. A finite transmission could be obtained here only for a very small barrier width $L$. The third transport process is the intraband process (between conduction bands), which appears for $\zeta > (U_0+\Delta)/E$. The intraband process induces a weaker valley contrast compared to the interband process, which can be understood by the dependence of the transmission with the parameters of the system.  

As can be seen, the transmittance for each valley is very sensitive to the Fermi velocity. A small change in $\zeta$ can induce a great change in the valley transmission through the barrier. As $B_0$ increases, the valley polarization improves, since the transmittance peaks of one valley match with a minimum of the other valley, as can be seen in the case with $B_0=3$, which does not occur for $B_0 = 1$. It can also be noted that the Fermi velocity modulation can be used to create confinement in graphene, since there is a range of values of $\zeta$ in which the transmittance is equal to zero for both valleys. This range does not change as the magnetic field increases, which means that, even for a weak magnetic field, the total reflection in the barrier can be achieved with the contribution of the Fermi velocity modulation. So, besides the improvement and control of the valley polarized current, the Fermi velocity can also turn on/off the electronic transport in graphene.

\begin{figure}[hpt]
\centering
\includegraphics[width=\linewidth]{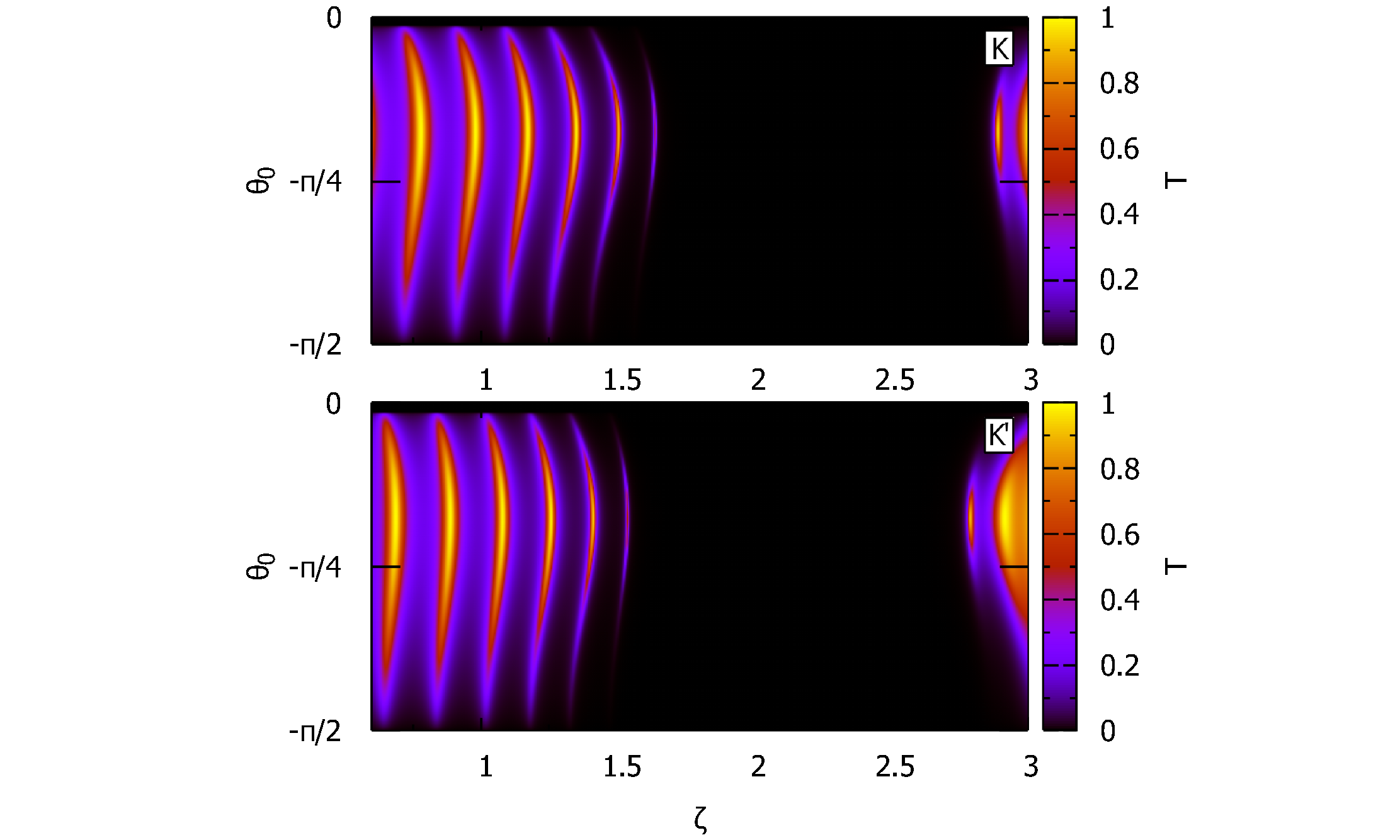}	
\caption{Contour plot of the transmittance as a function of incidence angle and $\zeta$ for both graphene valleys. We consider here $L=2$, $B=3$, $\Delta=4$, $E=7$ and $U_0 = 15.5$.}
\label{contour}
\end{figure} 

In Fig. \ref{E} we plotted the transmittance for different values of energy with a fixed magnetic field. In Figs. \ref{E} $(a)$, $(c)$ and $(e)$ we consider $T$ as a function of $k_y$ for various values of $\zeta$. As the energy increases, the range of $k_y$ with transmittance different from zero also increases, satisfying what was predicted by Eq. (\ref{sin}). The transmittance for both valleys becomes equal to one by choosing a specific value for $\zeta$, as can be seen in the dotted lines, which reinforce the relevance of a Fermi velocity modulation in the transport properties of graphene. So, besides of being able to make the filter totally reflect both valleys, a Fermi velocity modulation can also make it becomes transparent for all incidence angles included in the condition (\ref{sin}). In Figs. \ref{E} $(b)$, $(d)$ and $(f)$ we consider the transmittance as a function of $\zeta$ for $k_y=0$. One can see clearly again that the valley filter in very sensitive to a Fermi velocity modulation.  Also, it can be noted that, in contrast to the magnetic field, the range of values of $\zeta$ with $T=0$ for both valleys changes with the energy.

A contour plot of the transmittance as a function of 
$\zeta$ and incidence angle for valleys $K$ and $K^{\prime}$ can be seen in Fig. \ref{contour}, which reveals the influence of the Fermi velocity modulation for all incidence angles $\theta_0$. As can be seen, the transmittance for both valleys oscillates from 0 to 1 for almost all values of $\theta_0$ as $\zeta$ changes. Also, the peaks of transmittance for valley $K$ occur for different value of $\zeta$ than for the valley $K^{\prime}$, revealing that a Fermi velocity modulation can control the valley filter for quasiparticles in graphene with various incidence angles. 

\begin{figure}[hpt]
\centering
\includegraphics[width=0.9\linewidth]{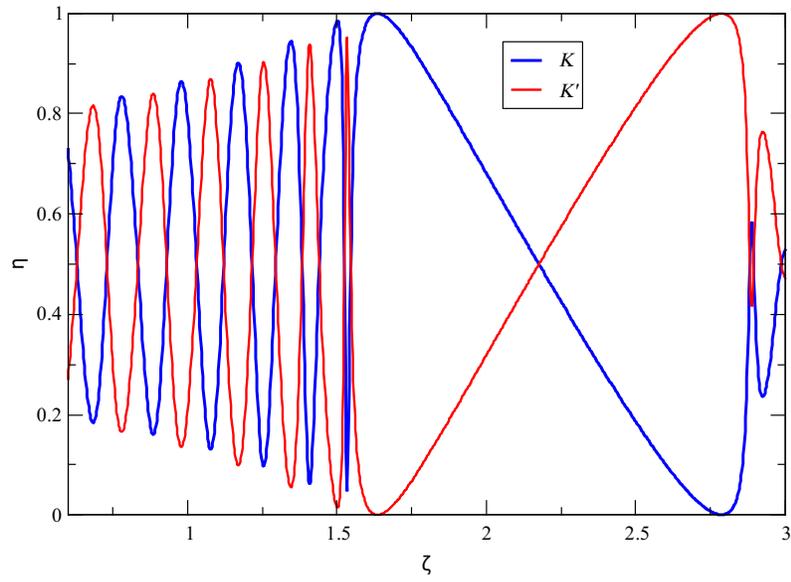}	
\caption{The efficiency of the valley filter as a function of $\zeta$. We consider here $L=2$, $B=3$, $\Delta=4$, $E=7$, $U_0 = 15.5$ and $k_y = 0$.}
\label{efficiency}
\end{figure}

In Fig. \ref{efficiency} we consider the efficiency of the valley filter. One can note that the Fermi velocity modulation can improve the efficiency of the valley filter, leading to a perfect filter,  with efficiency 1 for one valley and zero for the other. This result, together with the previous ones, reveals that a controllable Fermi velocity in graphene is a powerful mechanism for the fabrication of a valley filter, since it has a total control in its properties.  

\begin{figure}[hpt]
\centering
\includegraphics[width=0.9\linewidth]{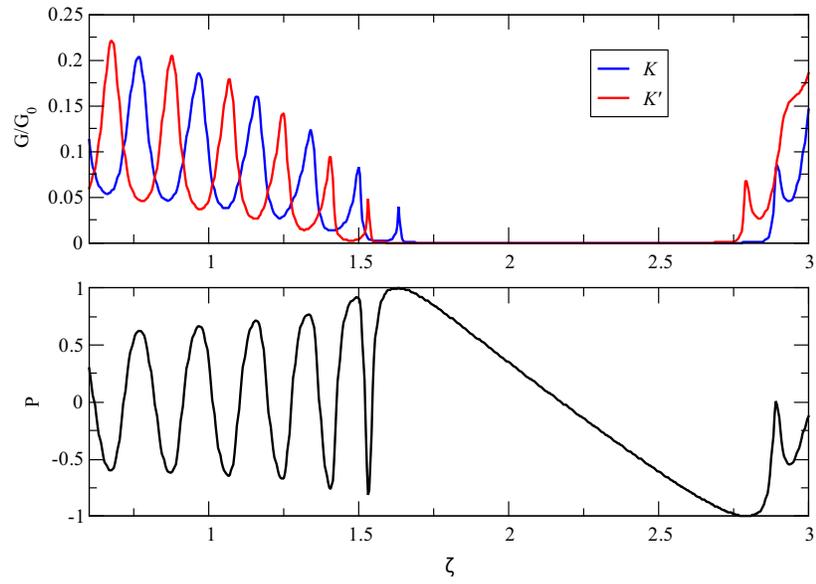}	
\caption{The conductance and the polarization of the valley filter as a function of $\zeta$. We consider here $L=2$, $B=3$, $\Delta=4$, $E=7$ and $U_0 = 15.5$.}
\label{cp}
\end{figure}

Finally, we calculated the conductance for each valley and the polarization of the filter as a function of the $\zeta$. It can be seen in Fig. \ref{cp}. Again, it is clear that the Fermi velocity modulation can improve and control the valley polarization in graphene. These results also reveal that this control is not restricted to a specific incidence angle, since to obtain the conductance we integrate the transmittance for all angles.	

\section{Conclusions}

In this paper we have demonstrated that a Fermi velocity modulation can improve the efficiency and control a valley filter in graphene. Considering that the filter is created by a combination of a magnetic and electric barrier, we showed that the Fermi velocity has a complete control in the valley polarization, being possible to choose which valley will be filtered and also to filter both or none of the valleys. Our results revealed that the valley filter is very sensitive to a modulation of the Fermi velocity, which means that it is a powerful tool to be used in a future graphene-based valleytronic device. 

{\bf Acknowledgements}: JRFL thanks I. Gornyi for discussions. This work was supported by CNPq, Capes and Alexander von Humboldt Foundation.


\end{document}